\documentclass[a4paper, oneside, twocolumn, notitlepage, 10pt]{extarticle_ecoc}
\usepackage{ecoc} 
\usepackage{pgfplots}
\pgfplotsset{compat=1.18}
\definecolor{chasered}{RGB}{169, 84, 84}%221,91,91
\definecolor{chasegreen}{RGB}{255, 90, 0}%95,181,243
\definecolor{chaseblue}{RGB}{100, 123, 177}%95,181,243
\definecolor{chasepurple}{RGB}{152, 96, 200}%168,122,231

\newsavebox{\singlechasebox}

\newsavebox{\rsbchbox}
\newsavebox{\ofecbox}
\newlength{\alignedplotheight}
\pgfdeclarelayer{background}
\pgfsetlayers{background,main}
\usetikzlibrary{calc}
\pgfdeclarelayer{background}
\pgfsetlayers{background,main}
\usetikzlibrary{calc}
\usepackage{subcaption}
\usepackage{xcolor}
\usepackage{bm}
\usepackage{balance}

\pgfdeclarelayer{background}
\pgfsetlayers{background,main}
\definecolor{customred}{rgb}{0.50, 0.00, 0.00}
\definecolor{customblue}{rgb}{0.00, 0.15, 0.50}
\begin{document}
\selectlanguage{english}    % Standard Language

%-------------------------------------------------- Title -----------------------------------------------------%

\title{Maximum Coverage Chase Decoder for Optical Interconnects}%

%------------------------------------------------- Authors-----------------------------------------------------%

\author{
    Alessandro~Cardinale\textsuperscript{(1)}, Wenqing~Song\textsuperscript{(1)},
    Bin~Chen\textsuperscript{(2)},
    Alex~Alvarado\textsuperscript{(3)},
    Andreas~Burg\textsuperscript{(1)}, Yifei~Shen\textsuperscript{(1)}
}

\maketitle                  % Create title and author

%------------------------------------------ Description of Authors ----------------------------------------------%

\begin{strip}
    \begin{author_descr}

        \textsuperscript{(1)} Telecommunications Circuits Laboratory, EPFL, Switzerland
        \textcolor{blue}{\uline{yifei.shen@epfl.ch}}

        \textsuperscript{(2)} School of Computer Science and Information Engineering, Hefei University of Technology, China

        \textsuperscript{(3)} Department of Electrical Engineering, Eindhoven University of Technology, The Netherlands

    \end{author_descr}
\end{strip}

% \setstretch{1.1}
%-------------------------------------------------- Footnote -------------------------------------------------------%
\renewcommand\footnotemark{}
\renewcommand\footnoterule{}
%\let\thefootnote\relax\footnotetext{text}

%-------------------------------------------------- Abstract ---------------------------------------------------------%
        % NOTE: Don't use a blank line here but start abstract right away to avoid an extra line break
        % Your 45-word abstract should be an explicit summary of the paper stating the problem, the methods used as well as the major results and conclusions. It should be complementary rather than a repetition of the title. Finish your abstract with the following copyright statement:

\begin{strip}
    \begin{ecoc_abstract}
        We propose a low-complexity Chase decoder for optical interconnects that formulates test pattern selection as a generalized maximum coverage problem. 
        For concatenated RS-BCH and oFEC codes, our decoder achieves the standard Chase decoding performance with 25\% and 61.3\% fewer test patterns, respectively. ©2026 The Author(s) 
    \end{ecoc_abstract}
\end{strip}

%-------------------------------------------------- Introduction Section -------------------------------------------------------%
\section{Introduction}
Next-generation optical interconnects~($1.6$~Tbps+) demand forward error correction~(FEC) that delivers high coding gain under stringent latency and power budgets.
Linear algebraic codes such as Bose-Chaudhuri-Hocquenghem~(BCH) and Reed–Solomon~(RS) codes, are widely employed due to their low-complexity hard-decision~(HD) decoders~\cite{wang_analysis_2017}.
With the adoption of higher-order modulation for improved spectral efficiency, soft-decision~(SD) decoding is used to obtain additional coding gain.
Among SD techniques for algebraic codes, the Chase algorithm~\cite{chase_class_1972} is the preferred choice, adopted in concatenated \mbox{RS-BCH} codes for short-reach links~\cite{matuz_serially_2025} and in open FEC~(oFEC) codes for medium-reach
links~\cite{mike_a_sluyski_open_2023} (Fig. \ref{fig:chase-diagram}).

\begin{figure}[t]
    \centering
    \setlength{\abovecaptionskip}{4.5pt}
    \pgfplotsset{
    label style={font=\fontsize{8pt}{7.2}\selectfont},
    tick label style={font=\fontsize{7pt}{7.2}\selectfont}
}

\begin{tikzpicture}
    \pgfplotsset{
            every axis/.append style={clip mode=individual}, 
            label style={font=\fontsize{8pt}{7.2}\selectfont},
            tick label style={font=\fontsize{7pt}{7.2}\selectfont}
    }
    \begin{axis}[
        name=topplot,
        xlabel={SNR [dB]},
        xlabel style={yshift=0.3em},
        ymode=log,
        ylabel={BER},
        ylabel style={yshift=-0.6em},
        xmin=14.3, xmax=16.5,
        ymin=1e-9, ymax=2e-6,
        grid=minor,
        ymajorgrids=true,
        xmajorgrids=true,
        grid style=dashed,
        width=0.95\linewidth,
        height=4.55cm,
        xtick={13.5, 14.0, 14.5, 15.0, 15.5, 16.0, 16.5},
        ytick={1e-9, 1e-8, 1e-7, 1e-6, 1e-5, 1e-4},
        yticklabels={$10^{-9}$, $10^{-8}$, $10^{-7}$, $10^{-6}$, $10^{-5}$, $10^{-4}$},
        axis background/.style={fill=white}
    ]
    \addplot[
        dashed,
        color=chasered,
        mark=*,
        mark size=2,
        every mark/.append style={solid},
        mark options={solid,fill=white},
        line width=1.2pt
    ]
    table {
        14.25  8.61506e-05
        14.50  3.23846e-05
        14.75  1.23807e-05
        15.00  4.00763e-06
        15.25  1.19956e-06
        15.50  3.31960e-07
        15.75  7.99584e-08
        16.00  1.82427e-08
        16.25  3.59277e-09
    };

    \addplot[
        dashed,
        color=chasegreen,
        mark=*,
        mark size=2,
        every mark/.append style={solid},
        mark options={solid,fill=white},
        line width=1.2pt
    ]
    table {
        14.25  7.73043e-05
        14.50  2.91917e-05
        14.75  1.08455e-05
        15.00  3.611848e-06
        15.25  1.01929e-06
        15.50  2.95735e-07
        15.75  7.62749e-08
        16.00  1.80102e-08
        16.25  3.42778e-09
    };

    \addplot[
        dashed,
        color=chaseblue,
        mark=*,
        mark size=2,
        every mark/.append style={solid},
        mark options={solid,fill=white},
        line width=1.2pt
    ]
    table {
        14.25  6.60460e-05
        14.50  2.54434e-05
        14.75  8.85669e-06
        15.00  3.01255e-06
        15.25  8.13618e-07
        15.50  2.20773e-07
        15.75  4.74609e-08
        16.00  1.03850e-08
        16.25  2.06531e-09
    };

    \addplot[
        color=chasered,
        mark=square*,
        mark size=2,
        every mark/.append style={solid},
        mark options={solid,fill=white},
        line width=1.2pt
    ]
    table {
        13.75  6.63336e-05
        14.00  2.26484e-05
        14.25  7.47376e-06
        14.50  2.11757e-06
        14.75  6.12828e-07
        15.00  1.26877e-07
        15.25  2.59949e-08
        15.50  4.27428e-09
    };

    \addplot[
        color=chasegreen,
        mark=square*,
        mark size=2,
        every mark/.append style={solid},
        mark options={solid,fill=white},
        line width=1.2pt
    ]
    table {
        13.75  7.53079e-05
        14.00  2.68534e-05
        14.25  8.71693e-06
        14.50  2.75249e-06
        14.75  8.30120e-07
        15.00  2.10076e-07
        15.25  4.75237e-08
        15.50  9.13274e-09
    };

    \addplot[
        color=chaseblue,
        mark=square*,
        mark size=2,
        every mark/.append style={solid},
        mark options={solid,fill=white},
        line width=1.2pt
    ]
    table {
        13.75  5.28730e-05
        14.00  1.79770e-05
        14.25  5.58667e-06
        14.50  1.37920e-06
        14.75  3.57863e-07
        15.00  6.61019e-08
        15.25  1.37021e-08
        15.50  1.94738e-09
    };

    % \node[anchor=north east, font=\bfseries] at (rel axis cs:0.02,0.95) {(a)};    
    \node[ anchor=north east, align=right, font=\fontsize{7.5pt}{7.2}\selectfont] at (rel axis cs:0.98,0.98) {$\bm{n=128}$ \\ $\bm{N_p=32}$};
    \end{axis}

    \begin{axis}[
        name=bottomplot,
        at={(topplot.south west)},
        anchor=north west,
        yshift=-0.8cm,
        xlabel={SNR [dB]},
        xlabel style={yshift=0.3em},
        ymode=log,
        ylabel={BER},
        ylabel style={yshift=-0.6em},
        xmin=15.0, xmax=16.85,
        ymin=1e-9, ymax=2e-6,
        grid=minor,
        ymajorgrids=true,
        xmajorgrids=true,
        grid style=dashed,
        width=0.95\linewidth,
        height=4.55cm,
        xtick={14.0, 14.5, 15.0, 15.5, 16.0, 16.5},
        ytick={1e-9, 1e-8, 1e-7, 1e-6, 1e-5, 1e-4},
        yticklabels={$10^{-9}$, $10^{-8}$, $10^{-7}$, $10^{-6}$, $10^{-5}$, $10^{-4}$},
        axis background/.style={fill=white}
    ]
    \addplot[
        dashed,
        color=chasered,
        mark=*,
        mark size=2,
        every mark/.append style={solid},
        mark options={solid,fill=white},
        line width=1.2pt
    ]
    table {
        15.00  5.17493e-05
        15.25  1.68562e-05
        15.50  4.98766e-06
        15.75  1.13655e-06
        16.00  2.37729e-07
        16.25  4.76148e-08
        16.50  9.28147e-09
        16.75  1.26530e-09
    };

    \addplot[
        dashed,
        color=chasegreen,
        mark=*,
        mark size=2,
        every mark/.append style={solid},
        mark options={solid,fill=white},
        line width=1.2pt
    ]
    table {
        15.00  4.45840e-05
        15.25  1.31569e-05
        15.50  3.75963e-06
        15.75  8.12596e-07
        16.00  1.71367e-07
        16.25  3.32313e-08
        16.50  6.28634e-09
        16.75  8.74629e-10
    };

    \addplot[
        dashed,
        color=chaseblue,
        mark=*,
        mark size=2,
        every mark/.append style={solid},
        mark options={solid,fill=white},
        line width=1.2pt
    ]
    table {
        15.00  3.82514e-05
        15.25  1.11964e-05
        15.50  2.89394e-06
        15.75  5.87932e-07
        16.00  1.21527e-07
        16.25  2.44168e-08
        16.50  4.32911e-09
        16.75  6.27179e-10
    };

    \addplot[
        color=chasered,
        mark=square*,
        mark size=2,
        every mark/.append style={solid},
        mark options={solid,fill=white},
        line width=1.2pt
    ]
    table {
        14.50  6.40016e-05
        14.75  1.95636e-05
        15.00  4.87535e-06
        15.25  9.77310e-07
        15.50  2.09425e-07
        15.75  3.29816e-08
        16.00  4.32529e-09
    };

    \addplot[
        color=chasegreen,
        mark=square*,
        mark size=2,
        every mark/.append style={solid},
        mark options={solid,fill=white},
        line width=1.2pt
    ]
    table {
        14.50  6.23322e-05
        14.75  1.90896e-05
        15.00  5.07333e-06
        15.25  1.16292e-06
        15.50  2.61266e-07
        15.75  4.34397e-08
        16.00  6.58885e-09
        %16.25  9.36011e-10
    };

    \addplot[
        color=chaseblue,
        mark=square*,
        mark size=2,
        every mark/.append style={solid},
        mark options={solid,fill=white},
        line width=1.2pt
    ]
    table {
        14.50  4.12389e-05
        14.75  1.12958e-05
        15.00  2.52757e-06
        15.25  5.14942e-07
        15.50  8.86575e-08
        15.75  1.20632e-08
        16.00  1.20152e-09
    };

    \node[anchor=north east, align=right, font=\fontsize{7.5pt}{7.2}\selectfont] at (rel axis cs:0.98,0.98) {$\bm{n=256}$ \\ $\bm{N_p=64}$};
    \end{axis}

    \node[anchor=north, draw=black, inner sep=1pt, xshift=0.35cm, yshift=0.02cm] at (current bounding box.south) {%
        \fontsize{7pt}{7.2}\selectfont
        \renewcommand{\arraystretch}{1.05}
        \begin{tabular}{@{}c@{}}
            \begin{tabular}{@{}c@{\hspace{8pt}}c@{}}
                \tikz[baseline=-0.6ex]{
                    \draw[dashed, color=black, line width=1pt] (0,0) -- (0.51,0);
                    \draw[color=black, line width=1.0pt, fill=white] (0.255,0) circle[radius=0.075];
                }\hspace{1pt}$t\!=\!2$ &
                \tikz[baseline=-0.6ex]{
                    \draw[color=black, line width=1pt] (0,0) -- (0.51,0);
                    \draw[color=black, line width=1.0pt, fill=white] (0.20,-0.065) rectangle (0.31,0.075);
                }\hspace{1pt}$t\!=\!3$
            \end{tabular} \\[-1.20ex]
            \tikz{\draw[dashed, color=black!70, line width=0.35pt] (-1,0) -- (2.55,0);} \\[0.25ex]
            \begin{tabular}{@{}p{1.5cm}@{}p{2.05cm}@{}p{2.05cm}@{}}
                \centering\tikz[baseline=-0.6ex]{\draw[color=chasered, line width=1.5pt] (0,0) -- (0.34,0);}\hspace{1pt}Chase-II &
                \centering\tikz[baseline=-0.6ex]{\draw[color=chasegreen, line width=1.5pt] (0,0) -- (0.34,0);}\hspace{1pt}LW-Chase &
                \centering\tikz[baseline=-0.6ex]{\draw[color=chaseblue, line width=1.5pt] (0,0) -- (0.34,0);}\hspace{1pt}GMC-Chase \tabularnewline
            \end{tabular}
        \end{tabular}
    };
\end{tikzpicture}%
    \caption{BER performance of single eBCH codes.}
    \label{fig:single-chase}
\end{figure}
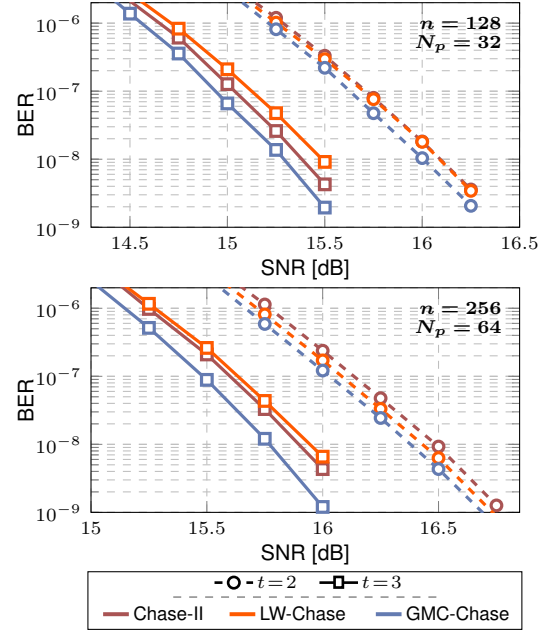

The Chase decoder flips unreliable bits based on test error patterns (TEPs) and performs HD decoding for each modified sequence.
Consequently, its decoding complexity depends heavily on the number of TEPs. A recent adaptive Chase decoder~\cite{yamamoto_simple_2026} reduces the average number of TEPs, but its worst-case complexity is identical to that of conventional Chase decoding~(\mbox{Chase-II}). \mbox{Alternative} methods focus on optimizing TEP selection.
Coded \mbox{Chase~\cite{tokushige_selection_2004, wu_novel_2023}} uses a short linear code to generate TEPs with large topological \mbox{coverage}. 
However, the generated TEPs are susceptible to erroneous bit flipping, making them unsuitable for high-rate codes~\cite{he_performance_2025}. 
{Inspired by~\cite{duffy_ordered_2022}}, a \mbox{logistic weight~(LW)} \mbox{metric} can be employed to generate TEPs.
LW-based Chase decoding (LW-Chase) is effective within the iterative oFEC framework~\cite{shen_iterative_2025}, but it yields marginal or negative gains over Chase-II for a single BCH code (Fig.~\ref{fig:single-chase}). Therefore, its benefit in concatenated RS-BCH schemes is limited because \mbox{single-code} decoding largely determines the overall performance. 

In this paper, we formulate TEP selection as a generalized maximum coverage (GMC) problem~\cite{cohen_generalized_2008} and propose a novel GMC-Chase \mbox{decoder}.
Our approach precomputes a fixed TEP set that maximizes the coverage of likely \mbox{error} patterns and penalizes patterns prone to erroneous bit flips. 
We demonstrate that GMC-Chase outperforms both Chase-II~\cite{chase_class_1972} and LW-Chase~\cite{shen_iterative_2025} on single BCH codes 
($2$- and $3$-bit error correction).
These gains translate into reduced complexity in concatenated RS-BCH and oFEC schemes, where the proposed GMC-Chase reduces the worst-case TEP number by $25\%$ and $61.3\%$ compared to the standard Chase decoder, respectively.

\section{Chase Decoding and TEP Generation}
\begin{figure}[t]
    \centering
    \includegraphics[width=1\linewidth]{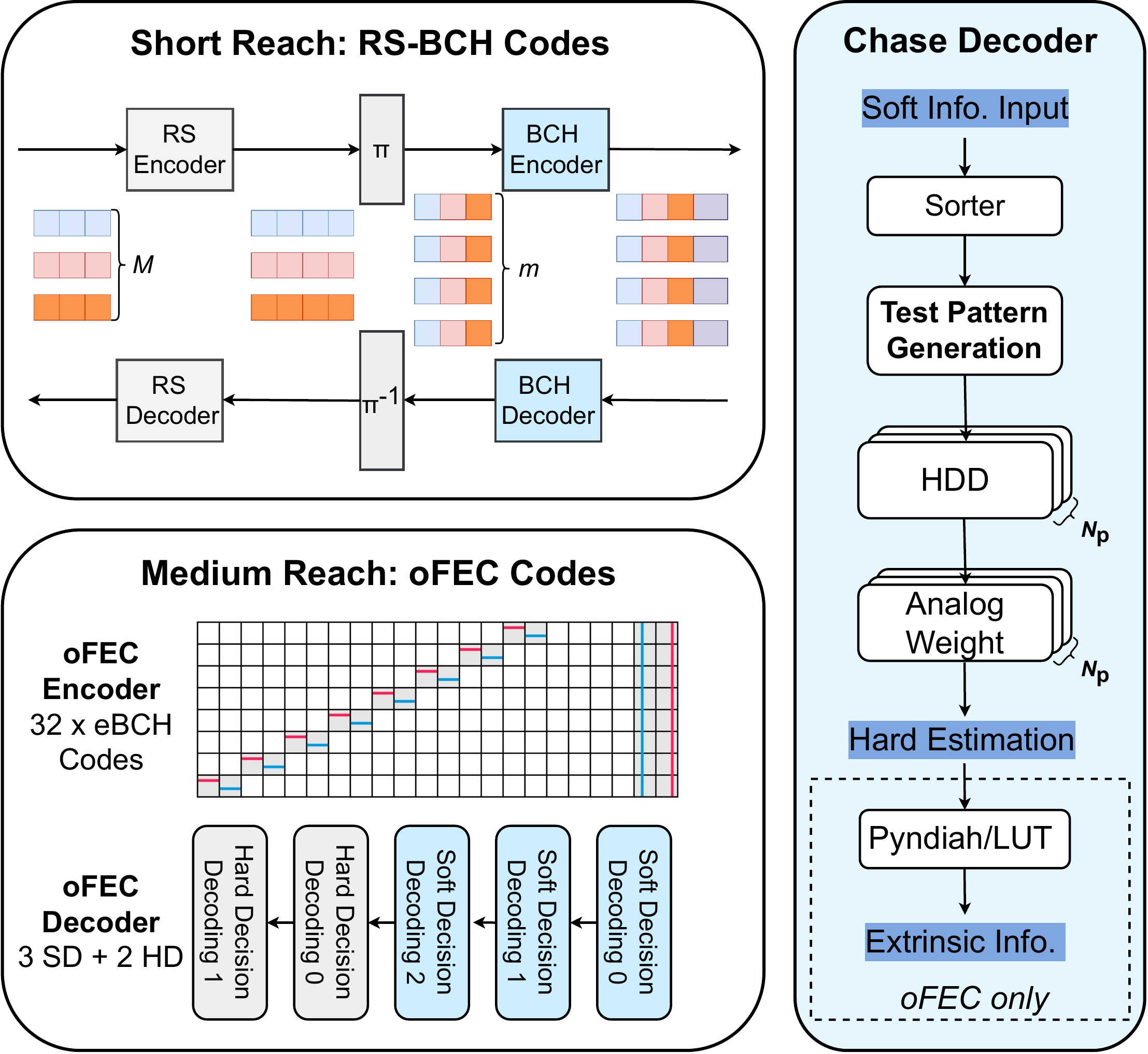}
    \caption{Chase decoder for short-reach RS-BCH codes and medium-reach oFEC codes.}
    \label{fig:chase-diagram}
\end{figure}
{Consider} a $\text{BCH}(n,k,t)$ code with error correction capability of $t$ bits and let $\boldsymbol{\lambda} = (\lambda_1, \dots, \lambda_n)$ denote the received bit log-likelihood ratios (LLRs). The corresponding HD sequence is denoted by~$\boldsymbol{r}$, with bit reliability measured by~$|\lambda_i|$.
In standard Chase-II decoding~\cite{chase_class_1972}, \mbox{$N_p = 2^J$} TEPs are generated by enumerating all binary combinations over the $J$ least reliable positions~(LRPs) in $\boldsymbol{r}$. 
Then, as depicted in Fig.~\ref{fig:chase-diagram}, for each TEP~$\boldsymbol{p}_q$, $q=1,2, \ldots, N_p$, the corresponding candidate sequence is formed as $\boldsymbol{r}_q = \boldsymbol{r} \oplus \boldsymbol{p}_q$ and is algebraically decoded into an estimated codeword~ $\boldsymbol{c}_q$. 
Among all candidate codewords, the Chase \mbox{decoder} outputs the one that minimizes the analog weight~\mbox{$W_q = \sum_{i=1}^{n} (r_i \oplus c_{q,i}) |\lambda_i|$}.
\begin{figure}[t]
    \centering
    \setlength{\abovecaptionskip}{8pt}
    \setlength{\belowcaptionskip}{0pt}
    % ==========================================
% SUBFIGURE A: Pure coverage
% ==========================================

\pgfplotsset{
    label style={font=\fontsize{8pt}{7.2}\selectfont},
    tick label style={font=\fontsize{7pt}{7.2}\selectfont}
}
\captionsetup[subfigure]{
    font=scriptsize,
    labelfont=bf,
    justification=centering,
    singlelinecheck=true,
    skip=1pt
}

\begin{subfigure}[t]{0.32\columnwidth} 
    \centering
    \resizebox{\linewidth}{!}{
        \begin{tikzpicture}[
            font=\fontsize{7.5pt}{8.4}\selectfont, 
            inner sep=0pt
        ]
            \useasboundingbox (-1.95, -2.12) rectangle (1.95, 1.95);

            \def\Rbig{1.8}
            \def\Rsmall{0.42} 
            \tikzset{dot/.style={circle, fill=black, inner sep=1pt}}
            \tikzset{leg_dot/.style={circle, inner sep=1.2pt}}

            \begin{pgfonlayer}{background}
                \fill[white] (0,0) circle (\Rbig);
                \draw[thick, gray!80!black] (0,0) circle (\Rbig);
                
                \foreach \x/\y in {0.8/0.8, -0.3/1.2, 0/0, -0.9/-0.7, -0.1/-1.2, 0.9/-0.8, 1.3/-0.1, -1.1/0.3} {
                    \fill[chaseblue, opacity=0.3] (\x,\y) circle (\Rsmall);
                    \draw[thin, chaseblue] (\x,\y) circle (\Rsmall);
                }
            \end{pgfonlayer}

            \node[dot, label={[label distance=0.5pt]above right:$\boldsymbol{p}_1$}] at (0.8,0.8) {};
            \node[dot, label={[label distance=0.5pt]above:$\boldsymbol{p}_2$}] at (-0.3,1.2) {};
            \node[dot, label={[label distance=0.5pt]right:$\boldsymbol{p}_3$}] at (0,0) {};
            \node[dot, label={[label distance=0.5pt]below left:$\boldsymbol{p}_4$}] at (-0.9,-0.7) {};
            \node[dot, label={[label distance=0.5pt]below:$\boldsymbol{p}_5$}] at (-0.1,-1.2) {};
            \node[dot, label={[label distance=0.5pt]below right:$\boldsymbol{p}_6$}] at (0.9,-0.8) {};
            \node[dot, label={[label distance=0.5pt]right:$\boldsymbol{p}_7$}] at (1.3,-0.1) {};
            \node[dot, label={[label distance=0.5pt]above right:$\boldsymbol{p}_8$}] at (-1.1,0.3) {};
        \end{tikzpicture}
    }
    % \vspace{0.4mm}
    \caption{Pure coverage}
    \label{subfig:pure_coverage}
\end{subfigure}
\hfill
% ==========================================
% SUBFIGURE B: Probability of error regions
% ==========================================
\begin{subfigure}[t]{0.32\columnwidth}
    \centering
    \resizebox{\linewidth}{!}{
        \begin{tikzpicture}[
            font=\fontsize{7.5pt}{8.4}\selectfont, 
            inner sep=0pt
        ]
            \useasboundingbox (-1.95, -2.12) rectangle (1.95, 1.95);

            % Parametri
            \def\Rbig{1.8}
            \def\Rmidlow{1.35}
            \def\Rmidhigh{0.9}
            \def\Rsmall{0.45}

            \begin{pgfonlayer}{background}
                \fill[gray!10] (0,0) circle (\Rbig);
                \fill[gray!25] (0,0) circle (\Rmidlow);
                \fill[gray!45] (0,0) circle (\Rmidhigh);
                \fill[gray!70] (0,0) circle (\Rsmall);
                \draw[thick, gray!80!black] (0,0) circle (\Rbig);
            \end{pgfonlayer}
        \end{tikzpicture}
    }
    % \vspace{0.4mm}
    \caption{Probability of error regions}
    \label{subfig:prob_regions}
\end{subfigure}
\hfill
% ==========================================
% SUBFIGURE C: Probability-aware coverage
% ==========================================
\begin{subfigure}[t]{0.32\columnwidth}
    \centering
    \resizebox{\linewidth}{!}{
        \begin{tikzpicture}[
            font=\fontsize{7.5pt}{8.4}\selectfont, 
            inner sep=0pt
        ]
            \useasboundingbox (-1.95, -2.12) rectangle (1.95, 1.95);

            \def\Rbig{1.8}
            \def\Rmidlow{1.35}
            \def\Rmidhigh{0.9}
            \def\Rsmall{0.45}
            \def\Rcover{0.42} 
            \tikzset{dot/.style={circle, fill=black, inner sep=1pt}}
            
            \begin{pgfonlayer}{background}
                \fill[gray!10] (0,0) circle (\Rbig);
                \fill[gray!25] (0,0) circle (\Rmidlow);
                \fill[gray!45] (0,0) circle (\Rmidhigh);
                \fill[gray!70] (0,0) circle (\Rsmall);
                \draw[thick, gray!80!black] (0,0) circle (\Rbig);
                
                \foreach \x/\y in {0/0, 0.5/0.5, -0.6/0.5, -0.3/-0.7, 0.5/-0.4, -0.7/-0.1, 1.1/0.0, -0.1/0.9} {
                    \fill[chaseblue, opacity=0.3] (\x,\y) circle (\Rcover);
                    \draw[thin, chaseblue] (\x,\y) circle (\Rcover);
                }
            \end{pgfonlayer}

            \node[dot, label={[label distance=0.5pt] left:$\boldsymbol{p}_1$}] at (0,0) {};
            \node[dot, label={[label distance=0.5pt]above right:$\boldsymbol{p}_2$}] at (0.5,0.5) {};
            \node[dot, label={[label distance=0.5pt]above left:$\boldsymbol{p}_3$}] at (-0.6,0.5) {};
            \node[dot, label={[label distance=0.5pt]below left:$\boldsymbol{p}_4$}] at (-0.3,-0.7) {};
            \node[dot, label={[label distance=0.5pt]below right:$\boldsymbol{p}_5$}] at (0.5,-0.4) {};
            \node[dot, label={[label distance=0.5pt]left:$\boldsymbol{p}_6$}] at (-0.7,-0.1) {};
            \node[dot, label={[label distance=0.5pt]right:$\boldsymbol{p}_7$}] at (1.1,0.0) {};
            \node[dot, label={[label distance=0.5pt]above:$\boldsymbol{p}_8$}] at (-0.1,0.9) {};
        \end{tikzpicture}
    }
    % \vspace{0.4mm}
    \caption{Probability-aware coverage}
    \label{subfig:prob_aware_coverage}
\end{subfigure}

\vspace{-0.1cm} 
\par\noindent
\begin{tikzpicture}[
    font=\fontsize{7pt}{8.4pt}\selectfont 
] 
    \tikzset{leg_dot/.style={circle, inner sep=1.8pt}}
    
    \node[leg_dot, fill=chaseblue, opacity=0.4] at (0,-0.38) {}; 
    \node[anchor=west] at (0.15,-0.38) {TEPs};

    \node[leg_dot, fill=gray!70] at (2.0,-0.38) {};
    \node[leg_dot, fill=gray!45] at (2.22,-0.38) {};
    \node[leg_dot, fill=gray!25] at (2.44,-0.38) {};
    \node[leg_dot, fill=gray!10] at (2.66,-0.38) {};
    \draw[->, thin, gray!80!black] (1.9,-0.54) -- (2.76,-0.54);
    \node[anchor=west] at (2.88,-0.38) {Lower probability};
\end{tikzpicture}
    \caption{Concept of improving error space coverage \\ by the TEPs.}
    \label{fig:overlap}
\end{figure}
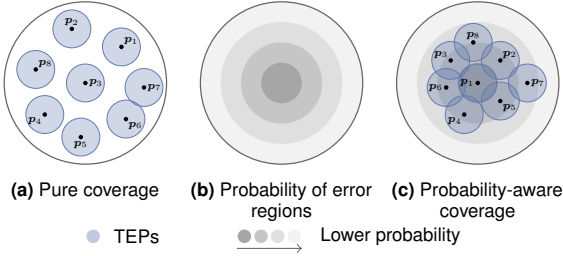

The LW metric~\cite{duffy_ordered_2022} provides an {alternative, reliability-based approach for generating TEPs}. 
{After sorting} bit positions in ascending order of reliability, {let $\tilde{\boldsymbol{e}}$ denote the permuted error pattern where $\tilde{e}_i$ refers to the $i$-th LRP.} The LW of $\tilde{\boldsymbol{e}}$ is defined as $\text{LW}(\tilde{\boldsymbol{e}}) = \sum_{i=1}^{n} i \cdot \tilde{e}_i$. 
Hence, smaller LW values prioritize TEPs concentrated on less reliable bit positions, which are statistically more likely to be erroneous.

\begin{figure}[t]
    \centering
    \setlength{\abovecaptionskip}{0pt}
    \input{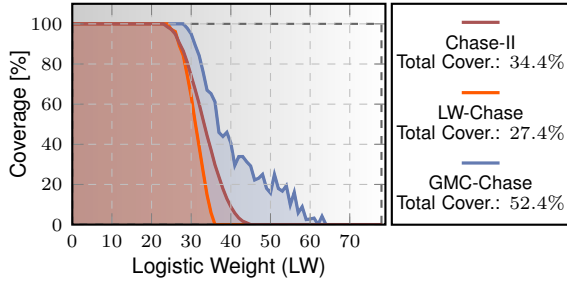}
    \caption{Example of error space coverage of Chase decoders ($\eta=12$, $N_p=64$,  $t=3$).}
    \label{fig:coverage}
\end{figure}

\section{Proposed GMC-based Test Pattern Selection}

A critical limitation of standard Chase decoding is TEP redundancy. 
Multiple candidate sequences often decode to the same codeword, which wastes computational resources. 

To improve decoding efficiency, the selected TEPs should be sufficiently diverse to correct a broad range of error patterns. This idea leads to the first core concept of our formulation: the coverage of the LRP error space (Fig.~\ref{fig:overlap}a). For a $\text{BCH}(n,k,t)$ code, a HD decoder guarantees the correction of up to $t$ errors. Assume a received sequence contains $E_{\text{LRP}}$ errors within the $\eta$ LRPs and $E_{\text{MRP}}$ errors in the most reliable positions (MRPs). Applying an $\eta$-bit TEP $\boldsymbol{p}$ to the LRPs introduces $F_c$ correct flips and $F_e$ erroneous flips. Successful HD decoding requires $E_{\text{LRP}} - F_c + F_e + E_{\text{MRP}} \leq t$. To account for residual errors outside the LRPs, we introduce a safety margin $\delta$ such that $E_{\text{MRP}} \leq \delta$. Thus, $\boldsymbol{p}$ can cover any LRP error sequence that satisfies $E_{\text{LRP}}-F_c+F_e \leq t - \delta$. So, $\boldsymbol{p}$ covers all error patterns within the Hamming ball $\mathcal{B}_{t-{\delta}}(\boldsymbol{p})$ of radius $t-{\delta}$ centered at $\boldsymbol{p}$.

The second observation is that error patterns are not equally likely (Fig.~\ref{fig:overlap}b).
Since at most $N_p$ TEPs can be tested, the decoder cannot cover the entire error space and must therefore allocate its TEPs diligently. The selected TEPs should preferentially cover the most probable error patterns.
In this work, the probability of a pattern is represented by its LW, where a lower LW indicates a higher probability.

A further point is that TEPs themselves are not equally desirable. Selecting an improbable TEP solely to cover a small additional error region is inefficient.
Therefore, effective TEPs should offer broad coverage of the error space, favoring the most probable error regions, and must also be highly probable themselves (Fig.~\ref{fig:overlap}c).
Because of this observation, we formulate TEP selection as a GMC problem~\cite{cohen_generalized_2008}, in which the profit of covering an error pattern depends on both the covered pattern and the TEP that covers it.

Let the universe $\mathcal{E}$ denote the set of all $2^\eta$ binary LRP error patterns $\tilde{\boldsymbol{e}}$. Each candidate TEP $\boldsymbol{p} \in \mathcal{E}$ induces a subset~$S_{\boldsymbol{p}}$ containing all error patterns in $\mathcal{E}$ within $\mathcal{B}_{t-{\delta}}(\boldsymbol{p})$. Selecting a subset~$S_{\boldsymbol{p}}$ is therefore equivalent to selecting
the TEP~$\boldsymbol{p}$. To set the constraint that at most $N_p$ TEPs are selected, we assign a unit weight to each subset, that is, $W(S_{\boldsymbol{p}})=1$ and zero weight to each $\tilde{\boldsymbol{e}}$ in~$S_{\boldsymbol{p}}$, namely, $W(\tilde{\boldsymbol{e}}, S_{\boldsymbol{p}})=0$.
To favor the coverage of statistically relevant error patterns by probable TEPs, we define the profit of covering $\tilde{\boldsymbol{e}}$ by
\begin{equation}
P(\tilde{\boldsymbol{e}},S_{\boldsymbol{p}}) = \frac{1}{\mathrm{LW}(\tilde{\boldsymbol{e}})\cdot\mathrm{LW}(\boldsymbol{p})}, \quad \forall\, \tilde{\boldsymbol{e}} \in S_{\boldsymbol{p}}.
\label{eq:profit}
\end{equation}
This definition assigns a large profit to the coverage of probable error patterns by probable TEPs, and a small profit to the coverage of improbable patterns or to coverage achieved using improbable TEPs. 
Consequently, the selection is biased \mbox{toward} a set of highly probable $N_p$~TEPs that jointly covers the most statistically relevant regions of the error space.
The GMC problem is solved using the greedy algorithm in~\cite{cohen_generalized_2008}, which yields the set of at most $N_p$ TEPs that maximizes the total profit accumulated over the covered error patterns.

\section{Numerical Results}

In the following case study, we set $\eta=12$, so that GMC-Chase selects
$N_p$ TEPs from a search space of $2^{12}$ LRP error patterns.
This choice provides a sufficiently rich candidate set for TEP selection.
Because the error probability decays rapidly across the MRPs, we set
$\delta=1$.

\vspace{0.17cm}
\textbf{\textit{Single eBCH Code}:}
We evaluate the proposed GMC-Chase decoder for single extended BCH (eBCH) codes over a four-level pulse amplitude modulation (PAM-4) system with Gray labeling.

Fig.~\ref{fig:coverage} visualizes the error space coverage for different Chase decoders with $N_p\!=\!64$ and $t\!=\!3$. Since LW-Chase strictly~minimizes LW, its coverage is clustered in the \mbox{low-LW} region, reaching only $27.4\%$ of the \mbox{total} patterns. Chase-II improves the coverage to~$34.4\%$ by incorporating some higher-weight patterns. Thanks to~(\ref{eq:profit}), \mbox{GMC-Chase} fully covers the critical \mbox{low-LW} region, while keeping a large distribution throughout the error space, achieving a $52.4\%$ total coverage. 
This broader coverage improves the error correction performance. As shown in Fig.~\ref{fig:single-chase}, with the same number of TEPs GMC-Chase consistently outperforms both \mbox{Chase-II} and \mbox{LW-Chase} across all tested cases: $\text{eBCH}(128,113,2)$ and $(128,106,3)$ with $N_p=32$, alongside $\text{eBCH}(256,239,2)$ and $(256,231,3)$ with $N_p=64$.

\vspace{0.17cm}
\textbf{\textit{RS-BCH Code}:}
\begin{figure*}[t]
    \centering
    \sbox{\rsbchbox}{%
        \begin{minipage}{0.49\textwidth}
            \centering
            \pgfplotsset{
    label style={font=\fontsize{8pt}{7.2}\selectfont},
    tick label style={font=\fontsize{7pt}{7.2}\selectfont}
}

\begin{tikzpicture}
    \begin{axis}[
        xlabel={Pre-FEC BER / SNR},
        xlabel style={align=center, yshift=0.2em},
        ymode=log,
        ylabel={Post-FEC BER},
        ylabel style={yshift=-0.5em},
        xmin=13.98, xmax=14.67,
        ymin=1e-15, ymax=1e-4,
        ytick={1e-15, 1e-14, 1e-13, 1e-12, 1e-11, 1e-10, 1e-9, 1e-8, 1e-7, 1e-6, 1e-5, 1e-4},
        yticklabels={$10^{-15}$, , $10^{-13}$, , $10^{-11}$, , $10^{-9}$, , $10^{-7}$, , $10^{-5}$, },
        minor ytick={2e-15, 3e-15, 4e-15, 5e-15, 6e-15, 7e-15, 8e-15, 9e-15, 2e-14, 3e-14, 4e-14, 5e-14, 6e-14, 7e-14, 8e-14, 9e-14, 2e-13, 3e-13, 4e-13, 5e-13, 6e-13, 7e-13, 8e-13, 9e-13, 2e-12, 3e-12, 4e-12, 5e-12, 6e-12, 7e-12, 8e-12, 9e-12, 2e-11, 3e-11, 4e-11, 5e-11, 6e-11, 7e-11, 8e-11, 9e-11, 2e-10, 3e-10, 4e-10, 5e-10, 6e-10, 7e-10, 8e-10, 9e-10, 2e-9, 3e-9, 4e-9, 5e-9, 6e-9, 7e-9, 8e-9, 9e-9, 2e-8, 3e-8, 4e-8, 5e-8, 6e-8, 7e-8, 8e-8, 9e-8, 2e-7, 3e-7, 4e-7, 5e-7, 6e-7, 7e-7, 8e-7, 9e-7, 2e-6, 3e-6, 4e-6, 5e-6, 6e-6, 7e-6, 8e-6, 9e-6, 2e-5, 3e-5, 4e-5, 5e-5, 6e-5, 7e-5, 8e-5, 9e-5},
        grid=minor,
        thick,
        xtick={14.0, 14.1, 14.2, 14.3, 14.4, 14.5, 14.6},
        xticklabels={
            {1.23\\[0.05cm]14.0}, 
            {1.15\\[0.05cm]14.1}, 
            {1.07\\[0.05cm]14.2}, 
            {1.00\\[0.05cm]14.3}, 
            {0.93\\[0.05cm]14.4}, 
            {0.86\\[0.05cm]14.5}, 
            {0.80\\[0.05cm]14.6}
        },
        extra x ticks={14.67},
        extra x tick labels={[\%]\\[0.05cm]{[dB]}},
        extra x tick style={
        xticklabel style={align=center},
        grid=none,
        tick style={draw=none}
        },
        xticklabel style={align=center},
        minor x tick num=1,
        ymajorgrids=true,
        yminorgrids=true,
        xmajorgrids=true,
        xminorgrids=true,
        grid style=dashed,
        width=1\linewidth,
        height=7.1cm,
        axis background/.style={fill=white},
        legend style={
            anchor=south west,
            at={(0.02,0.02)},
            cells={anchor=west},
            nodes={scale=0.95, transform shape},
            inner xsep=1.5pt,
            row sep=0.2mm,
            font=\fontsize{7pt}{7.2}\selectfont,
            fill=white,
            draw=black!35
        },
        legend columns=1
    ]

    \addplot[color=chasered, mark=*, mark size=2, every mark/.append style={solid}, mark options={solid,fill=white}, line width=1.2pt, forget plot] table {
        14 0.000287540856031128
        14.05 0.000104801556420233
        14.1 2.74680796981014e-05
        14.15 7.01300624960833e-06
        14.2 1.8831e-06
        14.25 3.3268e-07
        14.3 5.70089244680991e-08
        14.35 9.4241e-09
        14.4 9.8054e-10
    };

    \addplot[color=chasegreen, mark=*, mark size=2, every mark/.append style={solid}, mark options={solid,fill=white}, line width=1.2pt, forget plot] table {
        14 0.000146529182879377
        14.05 4.98910505836576e-05
        14.1 1.14932639852961e-05
        14.15 3.00573687838615e-06
        14.2 8.42172592079313e-07
        14.25 1.35829324836341e-07
        14.3 2.48388212173867e-08
        14.35 2.5914e-09
        14.4 3.0755e-10
    };

    \addplot[color=chaseblue, mark=*, mark size=2, every mark/.append style={solid}, mark options={solid,fill=white}, line width=1.2pt, forget plot] table {
        14 0.000100661478599222
        14.05 3.03968871595331e-05
        14.1 6.38601568152044e-06
        14.15 1.6434e-06
        14.2 2.83262390825954e-07
        14.25 4.3946e-08
        14.3 6.2101e-09
        14.35 7.2043e-10
        14.4 5.0584e-11
    };

    \addplot[color=chaseblue, mark=diamond*, mark size=3, every mark/.append style={solid}, mark options={solid,fill=white}, line width=1.2pt, forget plot] table {
        14 0.000266031128404669
        14.05 9.53929961089494e-05
        14.1 2.24194904483507e-05
        14.15 6.57547669642399e-06
        14.2 1.6206e-06
        14.25 2.7581e-07
        14.3 4.7097e-08
        14.35 6.3424e-09
        14.4 5.7121e-10
    };

    \addplot[color=chasered, dashed, line width=1.2pt, forget plot] table {
        14.4 7.17979542966901e-10
        14.45 7.6609520543966e-11
        14.5 6.95722214166906e-12
        14.55 5.65252100875305e-13
        14.6 4.13372613918014e-14
        14.65 2.66385040066599e-15
        14.7 1.5792208526513e-16
    };

    \addplot[color=chasegreen, dashed, line width=1.2pt, forget plot] table {
        14.4 2.75871990868172e-10
        14.45 2.75212973648682e-11
        14.5 2.39887022125523e-12
        14.55 1.79006727199267e-13
        14.6 1.20602199870476e-14
        14.65 7.64426718548945e-16
    };

    \addplot[color=chaseblue, dashed, line width=1.2pt, forget plot] table {
        14.4 4.81135093316742e-11
        14.45 4.18037004310366e-12
        14.5 3.10373994380181e-13
        14.55 2.06559533792768e-14
        14.6 1.21338226792225e-15
        14.65 6.16900243315475e-17
    };

    \addplot[color=chaseblue, dashed, line width=1.2pt, forget plot] table {
        14.4 5.1018944379364e-10
        14.45 5.25999615382613e-11
        14.5 4.52592018077444e-12
        14.55 3.53185208848963e-13
        14.6 2.46829385898981e-14
        14.65 1.45256384057407e-15
        14.7 7.90410218408469e-17
    };
    
    \node[anchor=north east, align=right, font=\fontsize{9pt}{7.2}\selectfont] at (rel axis cs:0.98,0.98) {\textbf{PAM-4}};

    \coordinate (diamondPt) at (14.4, 5.7121e-10);
    \coordinate (redPt) at (14.47, 0.2e-10);

    \path (diamondPt) -- (redPt) coordinate[midway] (ovalCenter);

    \draw[thick, color=blue!0!black] (ovalCenter) ellipse [x radius=0.2cm, y radius=0.1cm];

    \def\gap{0.1em}       
    \def\angle{60}      
    \def\length{0.77cm}   
    
    \coordinate (ovalEdge) at ($(ovalCenter) + (\angle:0.2cm and 0.1cm)$);

    \draw[->, thick, color=blue!0!black, >=stealth] 
        ($(ovalEdge) + (\angle:\gap)$) -- ($(ovalEdge) + (\angle:\gap+\length)$);

    \node[anchor=south, align=center, font=\fontsize{9pt}{7.2}\selectfont, yshift=-0.5ex, xshift=1ex, text=blue!0!black] 
        at ($(ovalEdge) + (\angle:\gap+\length)$) {\textbf{25\%}\\[0.5ex]\textbf{complexity}\\[0.5ex]\textbf{reduction}};
    % --------------------------------------

    \addlegendimage{line width=1.2pt,color=black,every mark/.append style={solid},mark options={solid,fill=white}}
    \addlegendentry{Simulation}
    \addlegendimage{line width=1.2pt,color=black,dashed}
    \addlegendentry{Union bound}
    \addlegendimage{line width=1.2pt,color=chasered,mark=*,every mark/.append style={solid},mark options={solid,fill=white,mark size=2.5}}
    \addlegendentry{Chase-II ($N_p=64$)}
    \addlegendimage{line width=1.2pt,color=chasegreen,mark=*,every mark/.append style={solid},mark options={solid,fill=white,mark size=2.5}}
    \addlegendentry{LW-Chase ($N_p=64$)}
    \addlegendimage{line width=1.2pt,color=chaseblue,mark=*,every mark/.append style={solid},mark options={solid,fill=white,mark size=2.5}}
    \addlegendentry{GMC-Chase ($N_p=64$)}
    \addlegendimage{line width=1.2pt,color=chaseblue,mark=diamond*,every mark/.append style={solid},mark options={solid,fill=white,mark size=3.3}}
    \addlegendentry{GMC-Chase ($N_p=48$)}
    \end{axis}
\end{tikzpicture}
        \end{minipage}%
    }%
    \sbox{\ofecbox}{%
        \begin{minipage}{0.49\textwidth}
            \centering
            \pgfplotsset{
    label style = {font=\fontsize{8pt}{7.2}\selectfont},
    tick label style = {font=\fontsize{7pt}{7.2}\selectfont}
}

\begin{tikzpicture}
    \begin{axis}[
        xlabel={Pre-FEC BER / SNR},
        xlabel style={yshift=0.2em},
        ymode=log,
        ylabel={Post-FEC BER},
        ylabel style={yshift=-0.8em},
        xmin=2.1, xmax=2.47,
        ymin=1e-10, ymax=1e-3,
        grid=minor,
        ymajorgrids=true,
        xmajorgrids=true,
        grid style=dashed,
        thick,
        width=1\linewidth,
        height=7.1cm,
        xtick={1.8, 1.9, 2.0, 2.1, 2.2, 2.3, 2.4, 2.5},
        xtick={2.5, 2.4, 2.3, 2.2, 2.1, 2.0, 1.9, 1.8},
        x dir=reverse,
        xticklabel style={align=center},
        xticklabels={
            {2.5\\[0.05cm]12.25},
            {2.4\\[0.05cm]12.34},
            {2.3\\[0.05cm]12.43},
            {2.2\\[0.05cm]12.52},
            {[\%]\\[0.05cm]{[dB]}},
            {2.0\\[0.05cm]12.71},
            {1.9\\[0.05cm]12.81},
            {1.8\\[0.05cm]12.91}
        },
        extra x ticks={1.85},
        extra x tick labels={[\%]\\[0.05cm]{[dB]}},
        extra x tick style={
            xticklabel style={align=center},
            grid=none,
            tick style={draw=none}
        },
        ytick={1e-10, 1e-9, 1e-8, 1e-7, 1e-6, 1e-5, 1e-4, 1e-3, 1e-2},
        yticklabels={$10^{-10}$, $10^{-9}$, $10^{-8}$, $10^{-7}$, $10^{-6}$, $10^{-5}$,$10^{-4}$, $10^{-3}$,$10^{-2}$},
        axis background/.style={fill=white},
        legend style={
            anchor=south west,
            at={(0.02,0.02)},
            nodes={scale=0.92, transform shape},
            cells={anchor=west},
            inner xsep=1.5pt,
            row sep=0.2mm,
            font=\fontsize{7pt}{7.2}\selectfont,
            fill=white,
            draw=black!35
        },
        legend columns=1,
        legend cell align={left}
    ]

   \addplot [
        line width=1.2pt, 
        color=chasered,
        mark=*,
        mark size=2,
        every mark/.append style={solid},
        mark options={solid,fill=white},
        forget plot
        ]
    table {%
        2.3172     1.8e-3
        2.2828     3e-04 
        2.2493    3e-5
        2.2093    1.08e-06
        2.1819     8e-8
        2.1544    3.6e-9
        2.1265    1.05e-10
        2.1005    1.02e-11
    };

    \addplot [
        line width=1.2pt, 
        color=chasegreen,
        mark=square*,
        mark size=2,
        every mark/.append style={solid},
        mark options={solid,fill=white},
        forget plot
        ]
    table {%
        2.4481    7.50366e-03
        2.4249    3.68294e-03
        2.4019    1.33718e-03
        2.3790    3.56934e-04
        2.3562    8.83388e-05
        2.3335    1.72076e-05
        2.3110    2.11159e-06
        2.2886    2.56386e-07
        2.2664    8.55632e-09
        2.2443    8.45272E-11
    };

    \addplot [
        line width=1.2pt, 
        color=chaseblue,
        mark=square*,
        mark size=2,
        every mark/.append style={solid},
        mark options={solid,fill=white},
        forget plot
        ]
    table {%
        2.4481    5.14038e-03
        2.4249    2.30652e-03
        2.4019    8.16612e-04
        2.3790    2.40928e-04
        2.3562    5.67724e-05
        2.3335    7.68000e-06
        2.3110    1.06708e-06
        2.2886    7.20642e-08
        2.2664    2.09051e-09
        2.2553    9.75485e-11
    };

    \addplot [
        line width=1.2pt, 
        color=chaseblue,
        mark=diamond*,
        mark size=3,
        every mark/.append style={solid},
        mark options={solid,fill=white},
        forget plot
        ]
    table {%
        2.3335    2.44412e-03
        2.3110    1.10815e-03
        2.2886    4.81755e-04
        2.2664    1.09707e-04
        2.2443    2.29992e-05
        2.2222    2.78343e-06
        2.2004    4.05673e-07
        2.1786    2.19728e-08
        2.1463    2.73531e-10
        2.1355    5.79732E-11
    };
    \node[anchor=north east, align=right, font=\fontsize{9pt}{7.2}\selectfont] at (rel axis cs:0.98,0.98) {\textbf{16-QAM}};
    
    \coordinate (diamondPt) at (2.2004, 4.05673e-07); 
    \coordinate (redPt) at (2.185, 0.9e-07);        

    \path (diamondPt) -- (redPt) coordinate[midway] (ovalCenter);

    \draw[thick, color=blue!0!black] (ovalCenter) ellipse [x radius=0.2cm, y radius=0.1cm];

    \def\gap{0.1em}     
    \def\angle{60}       
    \def\length{0.6cm}   

    \coordinate (ovalEdge) at ($(ovalCenter) + (\angle:0.2cm and 0.1cm)$);

    \draw[->, thick, >=stealth, color=blue!0!black] 
        ($(ovalEdge) + (\angle:\gap)$) -- ($(ovalEdge) + (\angle:\gap+\length)$);

    \node[anchor=south, align=center, font=\fontsize{9pt}{7.2}\selectfont, yshift=-0.2ex, xshift=1ex, color=blue!0!black] 
        at ($(ovalEdge) + (\angle:\gap+\length)$) {\textbf{61.3\%}\\[0.5ex]\textbf{complexity}\\[0.5ex]\textbf{reduction}};
    % --------------------------------------

    \addlegendimage{line width=1.2pt,color=chasered,mark=*,mark size=2.5,every mark/.append style={solid},mark options={solid,fill=white}}
    \addlegendentry{\shortstack[l]{Chase-Pyndiah\\($N_p=93$)~\cite{wang_real-time_2023}}}
    \addlegendimage{line width=1.2pt,color=chasegreen,mark=square*,mark size=2.5,every mark/.append style={solid},mark options={solid,fill=white}}
    \addlegendentry{\shortstack[l]{LW-Chase-LUT\\($N_p=93$)~\cite{shen_iterative_2025}}}
    \addlegendimage{line width=1.2pt,color=chaseblue,mark=square*,mark size=2.5,every mark/.append style={solid},mark options={solid,fill=white}}
    \addlegendentry{\shortstack[l]{GMC-Chase-LUT\\($N_p=93$)}}
    \addlegendimage{line width=1.2pt,color=chaseblue,mark=diamond*,mark size=3.7,every mark/.append style={solid},mark options={solid,fill=white}}
    \addlegendentry{\shortstack[l]{GMC-Chase-LUT\\($N_p=36$)}}
    \end{axis}
\end{tikzpicture}
        \end{minipage}%
    }%
    \setlength{\alignedplotheight}{\dimexpr\ht\rsbchbox+\dp\rsbchbox\relax}%
    \ifdim\dimexpr\ht\ofecbox+\dp\ofecbox\relax>\alignedplotheight
        \setlength{\alignedplotheight}{\dimexpr\ht\ofecbox+\dp\ofecbox\relax}%
    \fi
    \begin{minipage}[t]{0.49\textwidth}
        \vspace{0pt}
        \centering
        \parbox[t][\alignedplotheight][t]{\linewidth}{\centering\usebox{\rsbchbox}}
        \vspace{-0.5cm}
        \captionof{figure}{BER performance of Chase decoding for RS-BCH: $25\times \text{RS}(544,514,15)$ - $544 \times \text{eBCH}(142,125,2)$.}
        \label{fig:RSBCH}
    \end{minipage}\hfill
    \begin{minipage}[t]{0.49\textwidth}
        \vspace{0pt}
        \centering
        \parbox[t][\alignedplotheight][t]{\linewidth}{\centering\raisebox{-3.43cm}[0pt][0pt]{\usebox{\ofecbox}}}
        \vspace{-0.5cm}
        \captionof{figure}{BER performance of Chase decoding for oFEC consisting of $32 \times \text{eBCH}(256,239,2)$.}
        \label{fig:oFEC}
    \end{minipage}
\end{figure*}
Fig.~\ref{fig:RSBCH} evaluates GMC-Chase, LW-Chase, and Chase-II for the concatenated coding system of Fig.~\ref{fig:chase-diagram}. The scheme consists of $M\!=\!25$ outer $\text{RS}(544, 514, 15)$ (KP4) codes and \mbox{$m\!=\!544$} inner $\text{eBCH}(142,125,2)$ codes, connected via a symbol-wise interleaver $\pi$. The employed \mbox{PAM-4} with multilevel coding (MLC) scheme and the union bound approximation are consistent with the method described in~\cite{sukmadji_performance-complexity-latency_2025}.
The proposed GMC-Chase decoder matches the Chase-II performance using only \mbox{$N_p=48$}, which translates to a $25\%$ reduction in decoding complexity.

\vspace{0.17cm}
\textbf{\textit{oFEC Code}:}
Fig.~\ref{fig:oFEC} compares the GMC-Chase decoder with existing Chase-based decoders for the standard oFEC code with $16$ quadrature amplitude modulation (16-QAM). 
The oFEC code is a semi-infinite spatially coupled code formed by $32$ eBCH($256,239,2$) component codes, which is typically decoded by window decoding with three SD iterations and two HD iterations. Chase decoding is used in SD iterations, which generates a list of candidate codewords and then outputs extrinsic soft information using the Pyndiah method~(Fig.\ref{fig:chase-diagram}). 

As reported in~\cite{wang_real-time_2023}, $N_p=93$ TEPs are sufficient to achieve the target performance required by 800G ZR~\cite{noauthor_open_2023}. A previous work~\cite{shen_iterative_2025} introduced \mbox{LW-based} TEP generation and look-up table (LUT) for soft-output update. As shown in Fig.~\ref{fig:oFEC}, with $N_p=93$, LW-Chase-LUT achieves a post-FEC bit error rate (BER) of $10^{-10}$ at an input BER of $2.25\times 10^{-2}$. On top of this LUT-based architecture, the proposed GMC-Chase provides a further performance gain. In particular, with only $N_p=36$, GMC-Chase-LUT attains nearly the same performance as the Chase-Pyndiah baseline with $N_p=93$, which results in a $61.3\%$ reduction in computational complexity.

\vspace{1mm}
\section{Conclusions}
This work shows that the efficacy of TEPs in Chase decoding relies on both their statistical relevance and their coverage of probable error regions. To jointly optimize these two objectives, we formulate TEP selection as a GMC problem. For short- and medium-reach optical links, the proposed GMC-Chase decoder maintains the performance of the standard Chase decoder with lower complexity.

\clearpage
\section{Acknowledgements}
This work was supported in part by the Swiss State Secretariat for Education, Research, and Innovation (SERI) under the SwissChips initiative, in part by the National Natural Science Foundation of China under Grant No. 62571176, in part by the Anhui Provincial Natural Science Foundation under Grant No. 2508085J041, and in part by the BIT-FREE project with file number 20348, within the Open Technology Programme, which is partly financed by the Dutch Research Council (NWO).

\defbibnote{myprenote}{
}
\printbibliography[prenote=myprenote]

\vspace{-4mm}

\end{document}